\documentclass[11pt,fleqn]{article}
\usepackage{a4wide}
\newcommand{\req}[1]{(\ref{#1})}

\newcommand{\beq}{\begin{equation} }
\newcommand{\eeq}{\end{equation} }

\newcommand{\pa}{\partial}
\newcommand{\Res}{\mathop{\rm Res}\nolimits}

\newcommand{\papernumber}{ IFT/19/95 \\ hep-th/9511083 }

\title{Remarks on tree-level topological string theories}

\author{Robert J. Budzy\'{n}ski
	\thanks{email: {\tt budzynsk@fuw.edu.pl}}  \\
	{\it Institute of Theoretical Physics,
		Warsaw University } \\
		Ho\.{z}a 69, 00-681 Warsaw, Poland
	}

\date{\papernumber \\ November 13, 1995 }

\begin{document}

\maketitle

\abstract{A few observations concerning topological string theories at
the string-tree level are presented: (1) The tree-level, large phase
space solution of an arbitrary model is expressed in terms of a
variational problem, with an ``action'' equal, at the solution, to the
one-point function of the puncture operator, and found by solving
equations of Gauss-Manin type; (2) For $A_k$ Landau-Ginzburg models,
an extension to large phase space of the usual residue formula for
three-point functions is given.}
\thispagestyle{empty}
\newpage

\setcounter{page}{1}
\section{Introduction}

Topological string theory \cite{TopPhase} has been successful in
providing a rather simple and convenient language to describe the
minimal models of noncritical string theory \cite{DijkWitt,DVV}. There
has also been some progress in attempts to extend this language to
noncritical strings in two-dimensional target space, though this has
proved to be not quite straightforward
\cite{MuVa,GhoMu,IK,U,EguKa,ImMu}. The topological description is
especially efficient in reproducing the string tree level (genus zero)
approximation to noncritical string amplitudes.

A number of open questions remain, however, concerning the relevance
of the topological formulation to more physically interesting string
theory models. On one hand, it seems rather straightforward to couple
any topological matter theory, as originally defined by Dijkgraaf,
Verlinde and Verlinde (see \cite{LG} for an early review), to
topological gravity --- at least in the genus zero approximation, but
what remains unclear is whether consistency of such a construction can
in general be preserved at higher genera, or whether requiring this
imposes additional constraints on the topological matter theory,
beyond those laid down in \cite{LG}. In other words, we lack a well
defined and universal prescription to compute loop corrections to
topological string theory amplitudes from the data defining the theory
at the level of the spherical approximation, and thus we cannot test
the consistency and uniqueness of string loop corrections for a
generic model.

A second open question concerns the generalization of the topological
string theory formalism to models with an infinite number of
topological primary fields (of which $c=1$ strings appear to provide
an example), and to theories with (physical) fermionic degrees of
freedom. The former question was raised in \cite{EY}, and the
interesting example of a topological string theory based on the $CP^1$
topological matter theory was worked out in \cite{CP1}. As for the
latter question, existing studies of the $c=1$ theories do not yet
seem to point to any general approach to the issue of models with
infinite primary fields, while the topological formulation of models
with fermionic degrees of freedom has not been much studied; perhaps
further work using the results of \cite{A-Getal1,A-Getal2,B+B} will
provide clues in this direction.

In the present note, I wish to first point out a few simple properties
of tree-level topological string theory that follow rather directly
from its basic axioms, and therefore will hold for the coupling of any
topological matter model (with a finite number of topological primary
fields) to topological gravity. This is basically a reformulation and
extension of the discussion in Section 2 of \cite{EY}: namely, the
tree-level, large phase space solution of an arbitrary topological
string theory may be expressed in terms of a variational problem, with
an ``action'' (eq. \req{Vareq}) equal, at the solution, to the
one-point function of the puncture operator, and found by solving
differential equations of the Gauss-Manin type (eq. \req{GM}). It
should be stressed that the stated property holds independently of
whether or not a topological string theory admits a Landau-Ginzburg
realization; this fact has not been clearly stated in the literature,
and hopefully, it may be of some use in answering the open questions
discussed above. In the last section, I derive a simple generalization
(eq. \req{ResForm}) of the usual residue formula for three-point
functions of topological primaries in $A_k$ Landau-Ginzburg matter
theories, that is valid for general three-point correlators on large
phase space, and that follows straightforwardly from the integrable
(Gelfand-Dikii) structure of these theories, or --- more precisely ---
from its genus-zero (nondispersive) limit.

\section{The action for the string equations}

For the purposes of the present discussion, a tree-level topological
string theory is defined by the free energy of the underlying
topological matter system, i.e. a function $\cal F$ of the couplings
$t_{0,\alpha}$ to the topological primary fields $\Phi_{\alpha}$,
$\alpha = 0, \ldots, k$, whose derivatives with respect to the
couplings provide the correlators of primary fields on the ``small
phase space'' of the theory. $\cal F$ is required to satisfy the
constraints that
\beq
\langle P \Phi_{\alpha} \Phi_{\beta} \rangle \equiv
\frac{\pa^3\cal F}{\pa t_{0,0}\pa t_{0,\alpha}\pa t_{0,\beta}} =
\eta_{\alpha\beta} ,
\eeq
with $\eta_{\alpha\beta}$ a nondegenerate symmetric tensor independent
of the couplings (the topological metric), and that
\beq
c_{\alpha\beta}^{\gamma} \equiv \langle \Phi_{\alpha} \Phi_{\beta}
\Phi_{\mu} \rangle \eta^{\mu\gamma},
\eeq
(where $\eta^{\alpha\beta}$ is the inverse of $\eta_{\alpha\beta}$)
provide, for all values of the primary field couplings, a set of
structure constants for an associative, commutative algebra. That is,
one may define a law of multiplication of primary fields, by
\beq
\Phi_{\alpha} \cdot \Phi_{\beta} = c_{\alpha\beta}^{\gamma}
\Phi_{\gamma},
\eeq
such that
\beq
\Phi_{\alpha} \cdot \Phi_{\beta} = \Phi_{\beta} \cdot \Phi_{\alpha}
\eeq
and
\beq
(\Phi_{\alpha} \cdot \Phi_{\beta})\cdot \Phi_{\gamma} = \Phi_{\alpha}
\cdot (\Phi_{\beta}\cdot \Phi_{\gamma}) ,
\eeq
and a nondegenerate symmetric scalar product,
\beq
(\Phi_{\alpha},\Phi_{\beta}) = \eta_{\alpha\beta} ,
\eeq
obeying
\beq
(\Phi_{\alpha},\Phi_{\beta}\cdot \Phi_{\gamma}) = (\Phi_{\alpha}\cdot
\Phi_{\beta}, \Phi_{\gamma}) .
\eeq
The {\em puncture operator} $P\equiv \Phi_0$ is seen to be the unit
element of this primary field algebra.

Given a topological matter system, i.e. a solution $\cal F$ to the
above constraints, the full topological string theory (at string-tree
level) is understood to be given by the extension of $\cal F$ to a
function on the ``large phase space'' of all (primary and descendant)
couplings $t_{n,\alpha}$ ($n=0,1,\ldots \infty$), that reduces to the
small phase space free energy for $t_{n,\alpha}=0$ ($n>0$), and is
determined by Witten's tree-level factorization relation:
\beq
\label{Wfact}
\langle \sigma_{n+1}(\Phi_{\alpha}) X Y \rangle = \langle
\sigma_n(\Phi_{\alpha}) \Phi_{\beta} \rangle \langle \Phi^{\beta} X Y
\rangle ,
\eeq
for arbitrary (primary or descendant) fields $X$, $Y$ (and
$\sigma_0(\Phi_{\alpha}) \equiv \Phi_{\alpha}$), together with the
puncture equation
\beq
\label{PEq}
\langle P \rangle = \frac{1}{2} \sum_{\alpha,\beta}
t_{0,\alpha}t_{0,\beta}\eta_{\alpha\beta} + \sum_{n=0}^{\infty}
t_{n+1,\alpha} \langle \sigma_n(\Phi_{\alpha}) \rangle .
\eeq

It has been pointed out in \cite{DijkWitt} that, as a consequence of
the above properties, all two-point functions, as functions on large
phase space, depend on the couplings $t_{n,\alpha}$ only through the
$k+1$ variables
\beq
u_{\alpha} = \langle P \Phi_{\alpha} \rangle .
\eeq
In other words, the expressions of arbitrary two-point functions in
terms of $u^{\alpha}$ are a universal characteristic of a given model,
i.e. do not contain the couplings explicitly, and are known as {\em
constitutive relations}. Of special interest are the constitutive
relations
\beq
\label{ConstRel}
\langle \Phi_{\alpha} \sigma_n(\Phi_{\beta}) \rangle =
R_{\alpha;k,\beta}(u) .
\eeq
The form of these constitutive relations can be rather
straightforwardly determined from the topological matter system, by
using (derivatives of) the puncture equation \req{PEq}, and the
factorization equation \req{Wfact}, restricted to small phase space.
Inserting the relations \req{ConstRel} into the equations obtained by
taking the first derivatives of \req{PEq} with respect to the primary
couplings, one obtains
\beq
\label{gLG}
u_{\alpha} = \sum_{\beta}\eta_{\alpha\beta}t_{0,\beta} +
\sum_{\beta}\sum_{n=0}^{\infty} t_{n+1,\beta} R_{\alpha;k,\beta}(u) ,
\eeq
a set of $k+1$ equations for the $k+1$ unknowns $u_{\alpha}$ in terms
of the couplings $t_{n,\alpha}$, called the ``generalized
Landau-Ginzburg equations'' in \cite{DijkWitt}. These determine the
full large phase space solution of the topological string theory at
genus zero.

Let us now consider the expression
\beq
-\frac{1}{2}\eta^{\alpha\beta}u_{\alpha}u_{\beta} +
\sum_{\alpha} \sum_{n=0}^{\infty} t_{n,\alpha} \langle P
\sigma_n(\Phi_{\alpha})\rangle,
\eeq
where the two-point functions are understood as functions of
$u_{\alpha}$ given by the constitutive relations.

On small phase space,
\beq
\frac{\pa}{\pa t_{0,\beta}} \langle P \sigma_n(\Phi_{\alpha})
\rangle = \langle P \Phi_{\beta} \sigma_n(\Phi_{\alpha}) \rangle
= \langle \Phi_{\beta} \sigma_{n-1}(\Phi_{\alpha}) \rangle
\eeq
for $n \neq 0$, where I used the small phase space form of the
puncture equation. But, being a relation between two-point
functions, the above extends to large phase space via the
constitutive relations, as
\beq
\frac{\pa}{\pa u_{\beta}} \langle P \sigma_n(\Phi_{\alpha})
\rangle = \eta^{\beta\gamma} \langle \Phi_{\gamma}
\sigma_{n-1}(\Phi_{\alpha}) \rangle ,
\eeq
while, for $n=0$,
\beq
\frac{\pa}{\pa u_{\beta}} \langle P \Phi_{\alpha} \rangle =
\delta_{\alpha}^{\beta}.
\eeq

Therefore, the $u$-derivatives (at constant $t$'s) of
\beq
\label{Vareq}
\langle P \rangle = -\frac{1}{2}\eta^{\alpha\beta}u_{\alpha}u_{\beta} +
\sum_{\alpha} \sum_{k=0}^{\infty} t_{k,\alpha} \langle P
\sigma_k(\Phi_{\alpha})\rangle (u_.)
\eeq
yield the generalized Landau-Ginzburg equations \req{gLG}.
Furthermore, we can indeed show that the expression \req{Vareq} is
equal (up to an integration constant independent of all the couplings)
to $\langle P \rangle$: this follows by observing that its total
derivative with respect to any coupling $t_{n,\alpha}$ equals $\langle
P \sigma_n(\Phi_{\alpha}) \rangle$ at the solutions of the string
equations \req{gLG}.

At this point, it is useful to observe that
\beq
\frac{\pa}{\pa u_{\alpha}} \langle P \sigma_1(P) \rangle =
\eta^{\alpha\beta}u_{\beta},
\eeq
hence the first, quadratic term in \req{Vareq} may be absorbed into a
shift of the coupling $t_{1,0} \mapsto t_{1,0} -1$ (so that
$t_{1,0}=-1$ on small phase space).

In fact, a more general statement may be proven: namely, up to an
additive constant,
\beq
\label{Xeq}
\langle X \rangle = \sum_{\alpha}\sum_{n=0}^{\infty} t_{n,\alpha}
\langle X\sigma_n(\Phi_{\alpha}) \rangle ,
\eeq
for any operator $X$, primary or descendant. To see this, take the
derivative of the RHS of \req{Xeq} with respect to any of the
couplings $t_{n,\alpha}$, and use \req{Wfact} to show that the
contribution of terms where the derivative acts on the two-point
functions is proportional to the Landau-Ginzburg equations \req{gLG}.
Thus the derivatives of both sides of eq. \req{Xeq} with respect to
any of the couplings are equal, up to terms that vanish by the string
equations, proving our claim.

It should also be noted that eq. \req{Xeq} is nothing else than a
derivative of the (tree-level) {\em dilaton equation},
\beq
\sum_{\beta}\sum_{n=0}^{\infty} t_{n,\beta} \langle
\sigma_n(\Phi_{\alpha}) \rangle = 2{\cal F} ,
\eeq
which was not itself assumed in the above.

We have thus seen that the tree level solution of an arbitrary
topological string theory satisfying our axioms is determined from a
variational problem, by the extremum of the ``action'' given by eq.
\req{Vareq} with respect to the ``order parameters'' $u^{\alpha}$.
The form of eq. \req{Vareq} for a given model is determined by its
constitutive relations, and those can be worked out from the small
phase space solution, i.e. from the underlying topological matter
theory. More explicitly, the small phase space form of the puncture
equation,
\beq
\langle PP\sigma_{n+1}(\Phi_{\alpha}) \rangle = \langle
P\sigma_n(\Phi_{\alpha}) \rangle
\eeq
leads to the following recursion for the constitutive relations:
\beq
\label{rec}
\frac{\pa}{\pa u^0} \langle P\sigma_{n+1}(\Phi_{\alpha}) \rangle =
\langle P\sigma_n(\Phi_{\alpha}) \rangle ,
\eeq
while the small phase space equality
\begin{eqnarray}
\langle P \sigma_{n+2}(\Phi_{\alpha})\Phi_{\beta}\Phi_{\gamma}
\rangle & = & \langle
\sigma_{n+1}(\Phi_{\alpha})\Phi_{\beta}\Phi_{\gamma} \rangle
=  \nonumber \\
\langle \sigma_n(\Phi_{\alpha})\Phi_{\delta} \rangle \langle
\Phi^{\delta} \Phi_{\beta} \Phi_{\gamma} \rangle & = & \langle PP
\sigma_{n+2}(\Phi_{\alpha}) \Phi_{\delta} \rangle \langle
\Phi^{\delta}\Phi_{\beta}\Phi_{\gamma} \rangle \nonumber
\end{eqnarray}
may be rewritten as
\beq
\frac{\pa}{\pa t_{0,\beta}} \frac{\pa}{\pa t_{0,\gamma}} \langle
P \sigma_{n+2}(\Phi_{\alpha}) \rangle =
c_{\beta\gamma}^{\delta}(t) \frac{\pa}{\pa t_{0,\delta}}
\frac{\pa}{\pa t_{0,0}} \langle P \sigma_{n+2}(\Phi_{\alpha})
\rangle ,
\eeq
which leads to a constraint on constitutive relations, valid on large
phase space, when one substitutes $t_{0,\alpha} \mapsto u^{\alpha}$:
\beq
\label{GM}
\frac{\pa}{\pa u^{\beta}} \frac{\pa}{\pa u^{\gamma}} \langle
P \sigma_{n}(\Phi_{\alpha}) \rangle =
c_{\beta\gamma}^{\delta}(u) \frac{\pa}{\pa u^{\delta}}
\frac{\pa}{\pa u^0} \langle P \sigma_{n}(\Phi_{\alpha})
\rangle ,
\eeq
(it is easily verified by hand that this equation is valid also for
$k=0,~1$).

Equations \req{rec} and \req{GM}, together with the initial conditions
\beq
\langle P\sigma_0(\Phi_{\alpha}) \rangle \equiv \langle P\Phi_{\alpha}
\rangle = u_{\alpha}
\eeq
determine all the constitutive relations in question, i.e. determine
the dependence of the ``action'' of eq. \req{Vareq} on $u_{\alpha}$,
up to additional arguments that must be invoked to fix integration
constants in \req{rec}. The equations \req{GM} generalize the
Gauss-Manin equations, known from the restricted context of
Landau-Ginzburg topological matter theories \cite{LG-GM}; here, they
are seen to hold for arbitrary topological matter theories.

\section{Large phase space residue formula for $A_k$ models}

To fix notations, I begin by recalling the formulas that state the
defining properties of $A_k$ topological strings \cite{LG}. The
algebra of primary fields is given in terms of the superpotential
$W(X)$,
\beq
W(X) = \frac{1}{k+2} X^{k+2} + \sum_{i=0}^{k} g_i(t)X^i ,
\eeq
where $t=\{t_{0,\alpha}\}$, $(\alpha = 0,\ldots ,k)$ are the couplings
to the primary fields, as the algebra of polynomials in $X$ modulo the
relation
\beq
\frac{dW}{dX} = 0 .
\eeq
The topological metric is provided by the formula
\beq
\eta_{\alpha\beta} = \Res_X \left[
\frac{\Phi_{\alpha}(X)\Phi_{\beta}(X)}{ W'(X) }\right] ,
\eeq
where $\Res_X$ denotes the residue at infinity in $X$, i.e. the
coefficient of the $X^{-1}$ term in a Laurent series expansion for
large $X$. A preferred basis of primary fields is determined by
requiring $\eta_{\alpha\beta}$ to be constant, and
\beq
\Phi_{\alpha}(X) = X^{\alpha} + {\cal O}(X^{\alpha -2}) .
\eeq
The $\Phi_{\alpha}$ are then given by the ($t$-dependent) polynomials
\beq
\label{PrimF}
\Phi_{\alpha}(X) = \frac{1}{\alpha +1} \frac{d}{dX} [L^{\alpha
+1}(X)]_+ ,
\eeq
where $L(X) = [(k+2)W(X)]^{\frac{1}{k+2}}$, $L(X) = X + {\cal
O}(X^{-1})$, and $[ \cdots ]_+$ denotes the polynomial part of the
large-$X$ Laurent series expansion of the expression inside the
brackets. The relation between the coefficients $g_i$ in the
superpotential and the primary couplings $t_{0,\alpha}$ is determined
by the equations
\beq
\label{PrimFlow}
\frac{\pa W}{\pa t_{\alpha}} = \Phi_{\alpha}.
\eeq
Further on, the structure constants of the primary field algebra ---
and equivalently, the small phase space three-point functions of the
primary fields, are given by the well-known residue formula
\beq
\langle \Phi_{\alpha}\Phi_{\beta}\Phi_{\gamma} \rangle =
\eta_{\gamma\mu}c^{\mu}_{\alpha\beta} = \Res_X \left(
\frac{\Phi_{\alpha}\Phi_{\beta}\Phi_{\gamma} }{ W' } \right).
\eeq
Finally, the solution to eq. \req{PrimFlow} may be written as
\beq
\label{ordpar}
\eta_{\alpha\beta}t_{\beta} = t_{k-\alpha} = \frac{1}{\alpha +1}
\Res_X (L^{\alpha +1}) ,
\eeq
completing the brief review of the well-known solution of the $A_k$
topological matter theories at genus zero.

To extend the description of these theories to correlators of
descendant fields and large phase space, observe first that, since on
small phase space
\beq
t_{k-\alpha} = \eta_{\alpha\beta} = \langle P \Phi_{\alpha} \rangle ,
\eeq
we may write a large phase space version of \req{ordpar} as
\beq
\label{OrdPar}
u_{\alpha} = \langle P\Phi_{\alpha} \rangle = \frac{1}{\alpha +1}
\Res_X (L^{\alpha +1}) .
\eeq
Effectively, we are extending the notion of superpotential to large
phase space, by defining it to be the same function as on small phase
space, but with the replacement $t_{0,\alpha} \mapsto u^{\alpha}$.
The well-known small phase space formulas that express various
correlators in terms of $W$ (or $L$) are then easily converted into
constitutive relations, provided we take care to use formulas that
involve two-point functions only. And thus, two-point functions of
primary fields are given by \cite{LG}
\beq
\langle \Phi_{\alpha}\Phi_{\beta} \rangle = \frac{1}{\alpha +1}
\Res_X(L^{\alpha +1}\Phi_{\beta}) ,
\eeq
with $\Phi_{\beta}$ given in terms of $L$ by eq. \req{PrimF}, and the
basic constitutive relations involving descendants read \cite{LG-GM}
\beq
\langle P \sigma_n(\Phi_{\alpha}) \rangle = \frac{ \Res_X(L^{\alpha
+1+n(k+2)}) }{ \prod_{i=0}^{n}(\alpha +1+i(k+2)) }.
\eeq

It is now easy to write a large phase space residue formula for
three-point functions involving a single descendant field insertion.
Since two-point functions depend on all couplings only through the
$u^{\alpha}$, we have
\beq
\langle \Phi_{\alpha}\Phi_{\beta}\sigma_n(\Phi_{\gamma}) \rangle =
\frac{\pa}{\pa t_{n,\gamma}} \langle \Phi_{\alpha}\Phi_{\beta} \rangle
= \frac{\pa u^{\mu}}{\pa t_{n,\gamma}} \frac{\pa}{\pa u^{\mu}} \langle
\Phi_{\alpha}\Phi_{\beta} \rangle ,
\eeq
and therefore
\beq
\langle \Phi_{\alpha}\Phi_{\beta}\sigma_n(\Phi_{\gamma}) \rangle =
\frac{\pa u^{\mu}}{\pa t_{n,\gamma}} \Res_X \left(
\frac{\Phi_{\alpha}\Phi_{\beta}\Phi_{\mu} }{ W' } \right).
\eeq
On the other hand, as the superpotential $W$ is also a function only
of the $u^{\alpha}$, we have
\beq
\frac{\pa W}{\pa t_{n,\gamma}} = \frac{\pa u^{\mu}}{\pa t_{n,\gamma}}
\frac{\pa W}{\pa u^{\mu}} = \frac{\pa u^{\mu}}{\pa t_{n,\gamma}}
\Phi_{\mu}(X).
\eeq
By combining the two above equations we obtain
\beq
\label{resform}
\langle \Phi_{\alpha}\Phi_{\beta}\sigma_n(\Phi_{\gamma}) \rangle =
\Res_X \left( \frac{ \Phi_{\alpha}\Phi_{\beta}\frac{\pa W}{\pa
t_{n,\gamma}} }{ W' }\right),
\eeq
the simplest instance of the claimed large phase space residue formula.

To further generalize this result, we recall here that the dependence
of W on the (primary and descendant) couplings may be described by an
infinite family of commuting Hamiltonian flows:
\beq
\label{PB}
\frac{\pa W}{\pa t_{n,\alpha}} = \left\{ H_{n,\alpha} , W \right\} ,
\eeq
with a Poisson bracket given by
\beq
\{ F,G \} = \frac{\pa F}{\pa X}\frac{\pa G}{\pa t_{0,0}} -
\frac{\pa F}{\pa t_{0,0}}\frac{\pa G}{\pa X},
\eeq
and the Hamiltonians
\beq
H_{n,\alpha} = \frac{ [L^{\alpha +1+n(k+2)}]_+ }{ \prod_{i=0}^{n}(\alpha
+1+i(k+2)) } ;
\eeq
in our notation, $\Phi_{\alpha}(X) = \pa_X H_{0,\alpha}$. The above is
simply the genus-zero limit of the Gelfand-Dikii integrable hierarchy
structure of these models that holds to all genera \cite{Dijk}. Now,
since the second term of the Poisson bracket eq. \req{PB} is a
polynomial in $X$ times $\frac{\pa W}{\pa X}$, the residue formula
\req{resform} may be written as
\beq
\langle \Phi_{\alpha}\Phi_{\beta}\sigma_n(\Phi_{\gamma}) \rangle =
\Res_X \left( \frac{ H_{0,\alpha}'H_{0,\beta}'H_{n,\gamma}' }{ W' }
\frac{ \pa W }{ \pa t_{0,0} }\right) .
\eeq
The claim is that for general three-point functions on large phase
space, this formula generalizes to
\beq
\label{ResForm}
\langle
\sigma_l(\Phi_{\alpha})\sigma_m(\Phi_{\beta})\sigma_n(\Phi_{\gamma})
\rangle = \Res_X \left( \frac{ H_{l,\alpha}' H_{m,\beta}'
H_{n,\gamma}' }{ W' } \frac{ \pa W }{ \pa t_{0,0} } \right) .
\eeq

This formula may be verified by using twice Witten's factorization
equation \req{Wfact} to reduce the general three-point function to one
with a single descendant field, and exploiting the identity
\beq
\label{auxid}
\frac{ \pa W }{ \pa t_{n,\alpha} } = \langle
\sigma_{n-1}(\Phi_{\alpha})\Phi^{\mu} \rangle \frac{ \pa W }{ \pa
t_{0,\mu} } .
\eeq

Equation \req{auxid} is itself a consequence of the
validity of constitutive relations, and of the puncture equation:
namely,
\beq
\frac{ \pa W}{ \pa t_{n,\alpha} } = \langle
P\Phi_{\mu}\sigma_n(\Phi_{\alpha}) \rangle \Phi^{\mu}(X) ;
\eeq
reexpressing the three-point function on the RHS as a derivative
of a two-point function, and using the small phase space form of the
puncture equation, the above equals
\beq
\frac{\pa u^{\nu}}{\pa t_{0,\mu}} \left( \frac{\pa}{\pa u^{\nu}}
\langle P \sigma_n(\Phi_{\alpha}) \rangle \right) \Phi^{\mu}(X) =
\langle P \Phi^{\nu} \Phi_{\mu} \rangle \langle
\sigma_{n-1}(\Phi_{\alpha}) \Phi_{\nu} \rangle \Phi^{\mu}(X)~.
\eeq
Finally, the first and last factor on the LHS above combine to
$\frac{\pa W}{\pa t_{0,\nu}}$, completing the proof of eq.
\req{auxid}. It is now straightforward to demonstrate the validity of
eq. \req{ResForm}, as outlined above.

\end{document}